\documentclass[aps, pra, reprint, superscriptaddress,longbibliography]{revtex4-2}
\usepackage{lineno,hyperref}
\usepackage{eurosym}
\usepackage{amsmath,amssymb,graphicx}
\usepackage{color}

\modulolinenumbers[5]
\newcommand{\bitem}{\begin{itemize}}
\newcommand{\fitem}{\end{itemize}}
\newcommand{\beq}{\begin{equation}}
\newcommand{\eeq}{\end{equation}}
\newcommand{\beqa}{\begin{eqnarray}}
\newcommand{\eeqa}{\end{eqnarray}}

\begin{document}

\title{Universal quantum Otto heat machine based on the Dicke model}

\author{He-Guang Xu}
\affiliation{School of Physics, Dalian University of Technology, 116024 Dalian,  China}
\author{Jiasen Jin}
\email{jsjin@dlut.edu.cn}
\affiliation{School of Physics, Dalian University of Technology, 116024 Dalian,  China}
\author{G.D.M. Neto}
\email{gdmneto@zjnu.edu.cn}
\affiliation{Department of Physics, Zhejiang Normal University, Jinhua 321004, China}
\author{Norton G. de Almeida}
\address{Instituto de Física, Universidade Federal de Goiás, 74.001-970, Goiânia
- GO, Brazil}

\date{\today}

\begin{abstract}
In this paper we study a quantum Otto thermal machine where the working substance is composed of N identical qubits coupled to a single mode of a bosonic field, where the atoms and the field interact with a reservoir, as described by the so-called open Dicke model. By controlling the relevant and experimentally accessible parameters of the model we show that it is possible to build a universal quantum heat machine (UQHM) that can function as an engine, refrigerator, heater or accelerator. The heat and work exchanges are computed taking into account the growth of the number N of atoms as well as the coupling regimes characteristic of the Dicke model for several ratios of temperatures of the two thermal reservoirs. The analysis of quantum features such as entanglement and second-order correlation shows that these quantum resources do not affect either the efficiency or the performance of the UQHM based on the open Dicke Model. In addition, we show that the improvement in both efficiency and coefficient of performance of our UQHM occurs for regions around the critical value of the phase transition parameter of the model.
\end{abstract}


\maketitle

\section{Introduction}

Quantum thermodynamics~\cite{scovil1959three,alicki1979quantum,kosloff1984quantum,quan2007quantum,quan2009quantum,brunner2012virtual,Kosloff2013Quantum,2013Quantum,goold2016d,Sai2016Quantum,2015Perspective,2017The,deffner2019quantum}, which is described by the laws of quantum mechanics and thermodynamics, plays a fundamental role in understanding the transitions between various forms of energy and become a vibrant branch of modern research. A quantum thermal machine ~\cite{quan2007quantum,maruyama2009colloquium,scully2010quantum,huang2012effects,2005Autonomous,2009Thermal,goswami2013thermodynamics,kieu2004second,allahverdyan2005work,2012Quantum} is a quantum device to study the thermodynamic properties of quantum systems. In recent years, the study of thermal nanomachines has been driven by the great theoretical and experimental effort dedicated to the investigation of their properties in the quantum regime. Nowadays, there are many experimental platforms to explore QHE, such as trapped ion systems~\cite{abah2012single,rossnagel2014nanoscale,rossnagel2016single,maslennikov2019quantum}, optomechanics~\cite{zhang2014quantum,2014Theory}, ultracold atoms~\cite{2012Isolated,brantut2013thermoelectric}, {nuclear magnetic resonance (NMR) ~\cite{batalhao2014experimental,micadei2019reversing, de2019efficiency}, superconducting circuits~\cite{quan2006maxwell,niskanen2007information,pekola2010decoherence,koski2015chip,pekola2016maxwell}.



Among thermal machines, great interest has been devoted to cyclic thermal machines, both refrigerators and engines, operating in the quantum regime where energy exchanges can occur, for example, between the reservoir and just two levels of a single atom or between levels of a quantum harmonic oscillator. Several typical quantum cycles have been extensively studied, such as Carnot, Otto, and Stirling cycles~\cite{quan2007quantum,2017The,deffner2019quantum,kieu2004second,feldmann1996heat,2004Characteristics,rezek2006irreversible,tonner2006quantum,barrios2021light,he2009performance,thomas2017implications,2017Quantum0,2021Finite}. In this paper, we are only concerned with quantum Otto cycle. The performance of quantum Otto cycle depend strongly on the choice of the working substance. For example, recent studies show that with two temperatures fixed, the Otto cycle performed with fermionic substances can surpass the performance of the same cycle when performed with bosonic substances \cite{de2021two}. Regarding the Otto cycle there are several cases being considered, such as single spin systems~\cite{feldmann2000performance,geva1992quantum}, two-level atoms~\cite{2012Quantum}, coupled spin systems~\cite{2004Characteristics,huang2013special,huang2014special}, coupled spin-$3/2$~\cite{ ivanchenko2015quantum}, harmonic oscillators~\cite{abah2012single,myers2020bosons}, relativistic oscillators~\cite{myers2021quantum}, Bose-Einstein condensates~\cite{wang2009performance,myers2022boosting}, and light-matter systems described by the Jaynes-Cumming ~\cite{altintas2015rabi,song2016one,barrios2017role,mojaveri2021quantum} and quantum Rabi~\cite{jaynes1963comparison,rabi1936process,rabi1937space} models.

Despite the several studies on light-matter systems, there are very little works devoted to investigate a quantum Otto heat engine operating with multi-qubits interacting with a single cavity mode in the dressed picture and taking into account dissipation as well as the number of two-level atoms, as described by the open Dicke model (ODM). Over the past decades, the Dicke model has been theoretically studied in several contexts, as for example quantum phase transition~\cite{hepp1973superradiant,wang1973phase,gross1982superradiance,buchhold2013dicke,kirton2017suppressing,dalla2016dicke,larson2017some,chen2008numerically}, quantum entanglement~\cite{schneider2002entanglement,lambert2004entanglement,wolfe2014certifying}, chaos~\cite{emary2003quantum}, lasing~\cite{kirton2018superradiant} and quantum thermodynamics~\cite{fusco2016work}. According to the qubit-photon coupling ratio $\lambda/\omega$, where $\lambda $ is the coupling strength and $ \omega$ is the frequency of the cavity mode field, the ODM can be divided into different coupling regimes: weak and strong coupling regime ($\lambda/\omega<0.1$), ultrastrong coupling regime (USC) ($0.1\leq\lambda/\omega<1$), which was experimentally realized in a variety of quantum systems ~\cite{ciuti2005quantum,gunter2009sub,niemczyk2010circuit,forn2010observation,zazunov2003andreev,janvier2015coherent,schwartz2011reversible,hoffman2011dispersive,scalari2012ultrastrong}, and deep strong coupling regime (DSC) ($1\leq\lambda/\omega$). 

In this work, we study a quantum Otto heat machine (QOHM) operating under two thermal reservoirs  and  having as working substance N atoms and one mode of an electromagnetic field, as modelled by the ODM. We calculate the total work extracted and the amount of heat exchanged between the system and the reservoir and  both the efficiency of the engine and the coefficient of performance (COP) of the refrigerator by numerically solving the ODM using the extended bosonic coherent state approach and the dressed master equation, which is suitable for any coupling strength regime to describe the ODM dynamics ~\cite{ le2016fate,beaudoin2011dissipation}. As we will show, it is possible, by controlling the ODM parameters, to build a universal quantum thermal machine \cite{de2020universal} that, depending on the choice of parameters, can work either as an engine, or as a refrigerator, or as heater, or as an accelerator. Furthermore, our results indicate that it is not possible, for the model analyzed here, to use quantum resources to improve the engine efficiency or the refrigerator performance.

This paper is organized as follows. In Sec. II we introduce the open Dicke model and numerically solve it by using the extended bosonic coherent state approach. In Sec. III we present our model for a universal QOHM, having as the working substance N two-level atoms and one mode of the electromagnetic field, both atoms and field interacting with their respective reservoirs through the so-called open Dicke model. In Sec. IV, we study the roles of the qubit-mode coupling strength, the number N of qubits and the temperature ratio between the cold and hot thermal reservoirs on the amount of work and heat extractable as well as the impact on the engine efficiency and the refrigerator performance when varying the system parameters. Finally, in Sec. V we present our conclusions.


\section{THE MODEL}
The Hamiltonian describing the Dicke model consisting of a single bosonic field interacting with  $N$ identical two-level qubits is expressed as ($\hbar=1$)~\cite{dicke1954coherence,kirton2019introduction}
\begin{equation}~\label{h0}
\hat{H}_0={\omega_{0}}\hat{a}^{\dagger}\hat{a}+\Delta \hat{J}_{z}+\frac{2\lambda}{\sqrt{N}} (\hat{a}^{\dagger}+\hat{a})\hat{J}_{x},
\end{equation}
where $\omega_{0}$ and $\Delta$ are the frequencies of the single bosonic mode and qubits, respectively, $\lambda$ is the qubit-boson coupling strength, $\hat{a}^{\dagger}$($\hat{a}$) denotes the creation (annihilation) operator of the bosonic field, $\hat{J}_{x}=\frac{1}{2}(\hat{J}_{+}+\hat{J}_{-})$ and $\hat{J}_{z}$ are the pseudospin operators given by $\hat{J}_{{\pm}}=\text{\ensuremath{\sum}}_{i}^{N}\hat{\sigma}_{\pm}^{i},\hat{J}_{z}=\sum_{i}^{N}\hat{\sigma}_{z}^{i}$, with $\hat{\sigma}_\alpha~(\alpha=x,y,z)$ being the Pauli operators.
The pseudospin operators satisfy the commutation relation
$[\hat{J}_{+},\hat{J}_{-}]=2\hat{J}_{z}$ , $[\hat{J}_{z},\hat{J}_{\pm}]=\pm \hat{J}_{\pm}$. In this work,  we will consider the resonance condition $\omega_{0}=\Delta=\omega$.

Dicke model has numerically exact solution by using extended bosonic coherent state approach~\cite{chen2008numerically}.
For convenience of numerical solution, we first rotate the angular momentum operators with $\pi/2$ along the $y$-axis
$\hat{H_{s}}=\exp(i{\pi}\hat{J}_y/2)\hat{H}_0\exp(-i{\pi}\hat{J}_y/2)$, resulting in
\begin{eqnarray}
\hat{H_{s}}=\omega_{0} \hat{a}^{\dagger}\hat{a}-\frac{\Delta}{2}(\hat{J}_{+}+\hat{J}_{-})
+\frac{2\lambda}{\sqrt{N}}(\hat{a}^{\dagger}+\hat{a})\hat{J}_{z}.
\end{eqnarray}
For the two-level qubits, its basis can be spanned by the Dicke state $\{|j,m{\rangle},m=-j,-j+1,...,j-1,j$\} with $j=N/2$,
and the Hilbert space of the total system can be expressed in terms of
the basis $\{|\varphi_{m}{\rangle}_{b}\otimes|j,m{\rangle}\}$.
In the Dicke model, the excitation number $\hat{N}=\hat{a}^\dag{\hat{a}}+\hat{J}_z+N/2$ is not conserved.
Therefore, the truncation of the bosonic excitation number procedure has to
be applied in this system, especially in the strong qubit-boson coupling regime.
By considering the displacement transformation $\hat{A}_{m}=\hat{a}+g_{m}$ with $g_{m}=2\lambda m/\omega\sqrt{N}$
and taking the total system basis into the Schr\"{o}dinger equation, we obtain
\begin{eqnarray}~\label{eq1}
&&-\Delta j_{m}^{+}|\varphi_{m}{\rangle}_{b}|j,m+1{\rangle}-\Delta j_{m}^{-}|\varphi_{m}{\rangle}_{b}|j,m-1{\rangle}\nonumber\\
&&+\omega_{0}(\hat{A}_{m}^{\dagger}\hat{A}_{m}-g_{m}^{2})|\varphi_{m}{\rangle}_{b}|j,m{\rangle}
=E|\varphi_{m}{\rangle}_{b}|j,m{\rangle},
\end{eqnarray}
where $\hat{J}_{\pm}|j,m{\rangle}=j_{m}^{\pm}|j,m\pm 1{\rangle}$, with $j_{m}^{\pm}=\sqrt{j(j+1)-m(m\pm1)}$.
Next, we multiply Eq.~(\ref{eq1}) on the left by $\{{\langle}n,j|\}$, which results in
\begin{equation}
-\Delta j_{n}^{+}|\varphi_{n+1}{\rangle}_{b}-\Delta j_{n}^{-}|\varphi_{n-1}{\rangle}_{b}+\omega_{0}(\hat{A}_{n}^{\dagger}\hat{A}_{n}-g_{n}^{2})|\varphi_{n}{\rangle}_{b}=E|\varphi_{n}{\rangle}_{b},
\end{equation}
where $n=-j,-j+1,...,j$.
Furthermore, the bosonic state can be expanded as
\begin{eqnarray}
|\varphi_{m}{\rangle}_{b}
&=&\sum_{k=0}^{\textrm{N}_\textrm{tr}}\frac{1}{\sqrt{k!}}
c_{m,k}(\hat{A}_{m}^{\dagger})^{k}|0{\rangle}_{A_{m}}\nonumber\\
&=&\sum_{k=0}^{\textrm{N}_\textrm{tr}}\frac{1}{\sqrt{k!}}
c_{m,k}(\hat{a}^{\dagger}+g_{m})^{k}e^{-g_{m}
\hat{a}^{\dagger}-g_{m}^{2}/2}|0{\rangle}_{a},
\end{eqnarray}
where $\textrm{N}_\textrm{tr}$ is the truncation number of bosonic excitations.
Finally, we obtain the eigen-value equation
\begin{eqnarray}
&&\omega_{0}(l-g_{n}^{2})c_{n,l}-\Delta j_{n}^{+}\sum_{k=0}^{\textrm{N}_\textrm{tr}}
c_{n+1,kA_{n}}{\langle}l|k{\rangle}_{A_{n+1}}\nonumber\\
&&-\Delta j_{n}^{-}\sum_{k=0}^{\textrm{N}_\textrm{tr}}
c_{n-1,kA_{n}}{\langle}l|k{\rangle}_{A_{n-1}}=Ec_{n,l},~
\end{eqnarray}
where the coefficients are $_{A_{n}}{\langle}l|k{\rangle}_{A_{n-1}}=(-1)^{l}D_{l,k}$ and
$_{A_{n}}{\langle}l|k{\rangle}_{A_{n+1}}=(-1)^{k}D_{l,k}$,
with
\begin{eqnarray}
D_{l,k}=e^{-G^{2}/2}\sum_{r=0}^{\min[l,k]}
\frac{(-1)^{-r}\sqrt{l!k!}G^{l+k-2r}}{(l-r)!(k-r)!r!}, G=\frac{2\lambda}{\omega_{0}\sqrt{N}}.
\end{eqnarray}
In the following work, we select the maximum  truncation number $\textrm{N}_\textrm{tr}=50$, which is sufficient to give the convergent excited state energies with relative error less than
$10^{-5}$.
As is well known, in the thermodynamic limit $N\to\infty$ the Dicke model undergoes a transition from normal (ground-state with zero photonic and atomic excitations) to superradiant (ground-state with a macroscopic population) phase when  the qubit-boson coupling strength crosses the critical value $\lambda_c=\frac{1}{2}\sqrt{\omega_{0}\Delta\coth(\beta\omega_{0}/2)}$. The zero and finite temperature transitions belong to different classes of universality with this difference manifested, for example, in photon-atom entanglement, which diverges for $T=0$ and remains finite for $T \ne 0$. Moreover, when $N=1$ the Dicke model is reduced to the seminal quantum Rabi model~\cite{rabi1936process,rabi1937space}.

To help clarify the numerical results, we explore two limit regimes of our
model, $(i)$ the thermodynamic limit with $N\rightarrow \infty $ and $\sqrt{N}%
\lambda =$\emph{constant}, and $(ii)$ the deep-strong coupling regime with fixed $%
N$ and $\lambda \rightarrow \infty .$ Both regimes allow to derive a
diagonalizable effective Hamiltonian through the Holstein-Primakoff (HP)
representation of the angular momentum operators which maps the total spin
operators $\hat{J}_{\alpha }$ to a bosonic mode $\hat{b}$~.

To the case $(i)$, the quantization axis is $\mathit{\,}\hat{J}_{z}\mathit{\,}$%
\ and the HP transformation, $\hat{J}_{z}=(\hat{b}^{\dagger }\hat{b}-\frac{N%
}{2})$, $\hat{J}_{+}=\hat{b}^{\dagger }\sqrt{N-\hat{b}^{\dagger }\hat{b}}$, $%
\hat{J}_{-}=\sqrt{N-\hat{b}^{\dagger }\hat{b}}$ $\hat{b},$ leading to the
large $N$ limit ( $N\gg \langle \hat{b}^{\dagger }\hat{b}\rangle $)

\begin{equation}~\label{eq10}
	\hat{H}_{HP(N)}={\omega _{0}}\hat{a}^{\dagger }\hat{a}+\Delta \hat{b}%
	^{\dagger }\hat{b}+\lambda (\hat{a}^{\dagger }+\hat{a})(\hat{b}^{\dagger }+%
	\hat{b}).
\end{equation}

The above Hamiltonian can be diagonalized in the normal phase $\lambda \leq 
\sqrt{{\omega _{0}}\Delta }/2=\lambda _{c}$ to $\hat{H}_{NP}={\ \varepsilon }%
_{-}c_{-}^{\dagger }c_{-}+{\varepsilon }_{+}c_{+}^{\dagger }c_{+}\,\ $with
the the energies given by 
\begin{equation}
	({\varepsilon }_{\pm })^{2}=\frac{{\omega _{0}^{2}+}\Delta ^{2}}{2}\pm \frac{%
		1}{2}\sqrt{({\omega _{0}^{2}-}\Delta ^{2})^{2}+16\lambda ^{2}{\omega _{0}}%
		\Delta }.\text{ \  }
\end{equation} After a suitable displacement of the HP bosons, the super-radiant phase $%
\lambda >\lambda _{c}\,$ can be cast in a bilinear form and diagonalized with the normal mode frequencies  
\begin{equation}
	({\varepsilon }_{\pm })^{2}=\frac{{\omega _{0}^{2}\lambda }^{4}{+}\Delta
		^{2}\lambda _{c}^{4}}{2\lambda _{c}^{4}}\pm \frac{1}{2\lambda _{c}^{4}}\sqrt{%
		({\omega _{0}^{2}{\lambda }^{4}-}\Delta ^{2}\lambda _{c}^{4})^{2}+4{\omega
			_{0}^{2}}\Delta ^{2}\lambda _{c}^{8}}.
\end{equation}

$~$The proper quantization axis to case $(ii)$ is $\hat{J}_{x}\mathit{\,}$\
and the HP leading Hamiltonian term to large $\lambda $ ( $N\gg
\langle \hat{b}^{\dagger }\hat{b}\rangle $) and eigenvalues limit are

\begin{eqnarray}~\label{eq11}
	\hat{H}_{HP(\lambda )} &=&{\omega _{0}}\hat{a}^{\dagger }\hat{a}+\frac{%
		4N\lambda ^{2}}{{\omega _{0}}}\hat{b}^{\dagger }\hat{b}-N\lambda (\hat{a}%
	^{\dagger }+\hat{a})  \nonumber \\
	E_{mn} &=&m\frac{4N\lambda ^{2}}{{\omega _{0}}}+n{\omega _{0}-}\frac{%
		N^{2}\lambda ^{2}}{{\omega _{0}}}.
\end{eqnarray}

We note that, the eigenstates of $\hat{H}_{HP(\lambda )}$ are product states
of photons displaced Fock states and atomic states being $x$-polarized.
Furthermore, both limiting cases lead to decoupled quantum harmonic
oscillators that can be used to calculate the average energy analytically
for each stage of the thermodynamic cycle and hence the work and efficiency.
\section{QUANTUM OTTO CYCLE}

\begin{figure}[tbp]
\begin{center}
\includegraphics[scale=0.25]{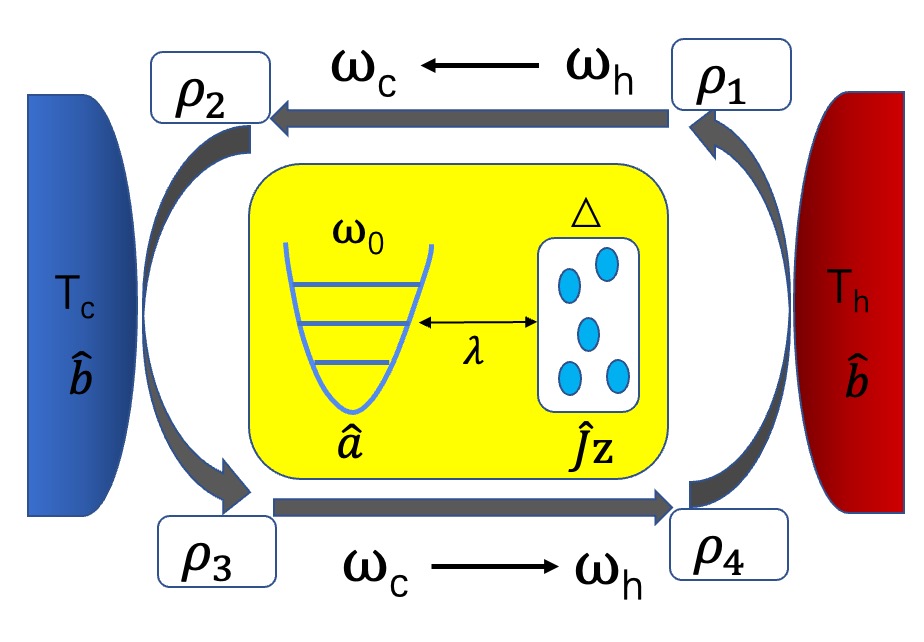}
\end{center}
\caption{Schematic representation of the four strokes of an Otto cycle for the realization of a universal heat machine based on the open Dicke mode, as detailed in Section III. During the isochoric stroke the frequency of the working substance, as modelled by the  Dicke Hamiltonian, is held fixed while  interacting with  a hot (cold) reservoir at temperature $T_{h}$  ($T_{c}$). Only heat is exchanged during this stroke. In the two quantum adiabatic strokes the working substance is isolated from the reservoir and has its frequency shifted, thus producing work. No heat is exchanged during this stroke. By controlling the parameters $\omega_{0}$, $\Delta$ and $\lambda$ of the model the machine can work as an engine, refrigerator, heater, or accelerator.
}~\label{fig001}
\end{figure}

To perform a quantum Qtto cycle, which is composed of two adiabatic and two isochoric processes~\cite{quan2007quantum,quan2009quantum}, we consider N two-level atoms and one electromagnetic field mode, as described by the Dicke model, as the working substance. During the isochoric process we left the N atoms and the electromagnetic field to interact with a hot (cold) reservoir at temperature $T_{h}$ ($T_{c}$ ). The four strokes of the quantum Otto cycle are described as follows (Fig.~\ref{fig001}).

1. Quantum isochoric process. The working substance as modelled by the  Dicke Hamiltonian
$H_{s}^{h}$ with frequency $\omega=\omega_{h}$ 
is brought into contact with a hot reservoir at temperature $T_{h}$. In this process,
the system undergoes a Markovian evolution, which is described by the quantum
dressed master equation~\cite{le2016fate,beaudoin2011dissipation} 
\begin{eqnarray}
\frac{d}{dt}{\hat{\rho}}_s&=&-i[{\hat{H}}_0,{\hat{\rho}}_s]+\sum_{u;k<j}
\{\Gamma^{jk}_un_u(\Delta_{jk})\mathcal{D}[|\phi_j{\rangle}{\langle}\phi_k|,{\hat{\rho}}_s]\nonumber\\
&&+\Gamma^{jk}_u[1+n_u(\Delta_{jk})]\mathcal{D}[|\phi_k{\rangle}{\langle}\phi_j|,{\hat{\rho}}_s]\},
\end{eqnarray}
where $|\phi_k{\rangle}$ is the dressed eigenbasis of the Dicke Hamiltonian $\hat{H}$ as $\hat{H}_0|\phi_k{\rangle}=E_k|\phi_k{\rangle}$, 
$\mathcal{{D}}[{\hat{O}},{\hat{\rho}}_s]
=\frac{1}{2}[2{\hat{O}}{\hat{\rho}}_s{\hat{O}}^{\dag}
-{\hat{\rho}}_s{\hat{O}}^{\dag}{\hat{O}}
-{\hat{O}}^{\dag}{\hat{O}}{\hat{\rho}}_s]$ is the dissipator,
$\Gamma^{jk}_u=\gamma_u(\Delta_{jk})|S^{jk}_u|^2$ is the rate,
with $S^{jk}_q=\frac{1}{\sqrt{N}}{\langle}\phi_j|({\hat{J}}_+
+{\hat{J}}_-)|\phi_k{\rangle}$
and $S^{jk}_c={\langle}\phi_j|({\hat{a}}^{\dag}+{\hat{a}})|\phi_k{\rangle}$,
where we consider the Ohmic case $\gamma_u(\Delta_{jk})=\pi\alpha\Delta_{jk}){\rm exp}(-|\Delta_{jk}|/\omega_{co})$, with $\alpha$ being the coupling strength and $\omega_{co}$ being the cutoff frequency of the thermal baths.
In the eigenbasis, the dynamics of the population $P_{n}=\langle \phi_{n}|\hat{\rho}_{s}|\phi_{n}\rangle$  is given by
\begin{eqnarray}
\frac{d}{dt}P_{n}
&=&\sum_{u,k{\neq}n}\Gamma^{nk}_un_u(\Delta_{nk})P_{k}\nonumber\\
&&-\sum_{u,k{\neq}n}\Gamma^{nk}_u[1+n_u(\Delta_{nk})]P_{n},
\end{eqnarray}
where $\Gamma^{nk}_u=-\Gamma^{kn}_u$. 

After a long enough evolution, the system will reach the only steady
state $\rho_{1}=\text{\ensuremath{\rho_{ss}}}(T_{h})=\sum_{n}P_{n}^{ss}(T_{h})|E_{n}^{h}\rangle\langle E_{n}^{h}|$ of Eq. (8) with
$\frac{d\rho}{dt}=0$, $P_{n}^{ss}(T_{h})$ being the corresponding population. The system eigenstates $|\phi_{k}^{h}\rangle$ and eigenvalues
$E_{k}^{h}$ of $H_{s}^{h}$ were obtained by using the extended bosonic
coherent state approach method ~\cite{chen2008numerically}. During this process, a heat amount $Q_{h}$ is absorbed from the hot
reservoir, without any work being done.

2. Quantum adiabatic expansion process. The system is isolated from
the hot reservoir and the energy levels is changed from $E_{n}^{h}$
to $E_{n}^{c}$ by varying the frequency from $\omega_{c}$ to $\omega_{h}$
(with $\omega_{h}>\omega_{c}$). This process must be done slow enough
to ensure that the populations $P_{n}^{ss}$ ($T_{h}$) remain unchanged
according to the quantum adiabatic theorem. At the end of this adiabatic expansion the state becomes $\rho_{2}=\sum_{n}P_{n}^{ss}(T_{h})|E_{n}^{c}\rangle\langle E_{n}^{c}|$ . During this process only
work is performed, with no heat being exchanged. 

3. Quantum isochoric process. The working substance
with frequency $\omega=\omega_{c}$ and modelled by the Hamiltonian $H_{s}^{c}$ is now put into contact with a cold reservoir at temperature $T_{c}<T_{h}$ until they reach thermal equilibrium. In this case, we have a change
in the steady state population from $P_{n}^{ss}$($T_{h}$) to $P_{n}^{ss}$($T_{c}$), while the eigenvalues $E_{n}^{c}$ of the system
 remain unchanged, and the state becomes $\rho_{3}=\sum_{n}P_{n}^{ss}(T_{c})|E_{n}^{c}\rangle\langle E_{n}^{c}|$. During this process, only heat is exchanged, an amount of
heat $Q_{c}$ is released to the reservoir, but no work is done. 

4. Quantum adiabatic compression process. The system is isolated
from the cold reservoir and its energy levels is changed back from
$E_{n}^{c}$ to $E_{n}^{h}$ by varying the frequency from $\omega_{h}$ to
$\omega_{c}$. At the end of the process, the populations $P_{n}^{ss}$
($T_{c}$) remain unchanged, the state becomes $\rho_{4}=\sum_{n}P_{n}^{ss}(T_{c})|E_{n}^{h}\rangle\langle E_{n}^{h}|$, and only work is performed on the working
substance, but no heat is exchanged. 

Next, let us calculate the work and heat exchanged in each stroke. According to the first law of thermodynamics, a quantum system with
discrete energy levels can be written as 
\begin{eqnarray}
dU=\delta Q+\delta W=\sum_{n}(E_{n}dP_{n}^{ss}+P_{n}^{ss}dE_{n}),
\end{eqnarray}
where $E_{n}$ are the energy levels and $P_{n}^{ss}$ are the occupation
probabilities at steady state.
Accordingly, the heat $Q_{h}$ ($Q_{c}$) exchanged with the hot (cold) reservoir, and the net work $W$ satisfy the following  relations~\cite{kieu2004second}
\begin{eqnarray}
Q_{h}=\sum_{n}E_{n}^{h}[P_{n}^{ss}(T_{h})-P_{n}^{ss}(T_{c})],
\end{eqnarray}
\begin{eqnarray}
Q_{c}=\sum_{n}E_{n}^{c}[P_{n}^{ss}(T_{c})-P_{n}^{ss}(T_{h})],
\end{eqnarray}
\begin{eqnarray}
W=Q_{h}+Q_{c}=\sum_{n}(E_{n}^{h}-E_{n}^{c})[P_{n}^{ss}(T_{h})-P_{n}^{ss}(T_{c})].
\end{eqnarray}
In this work we will adopt the following convention: $Q>0$ ($Q<0$) correspond to absorption (release) of heat from (to)
the reservoir while $W>0$ ($W<0$) correspond to work performed by
(on) the quantum heat engine. There are only four working regimes
allowed under not violating the Clausius inequality with the first
law of thermodynamics~\cite{solfanelli2020nonadiabatic}: (1) Heat engine (E): $Q_{h}>0$, $Q_{c}<0$,
and $W>0$; (2) Refrigerator (R): $Q_{c}>0$, $Q_{h}<0$, and $W<0$;
(3) Heater (H): $Q_{c}<0$, $Q_{h}<0$, and $W<0$; (4) Accelerator (A): $Q_{c}<0$,
$Q_{h}>0$, and $W<0$. In this article we are more concerned with the
heat engine and the refrigerator, which are of most interest for useful applications and whose figures of merit are the efficiency
$\eta=\frac{W}{Q_{h}}$ and the coefficient of performance (COP) $\xi=\frac{Q_{c}}{|W|}$,
respectively.

\section{Results and Discussions}

\begin{figure}[tbp]
\begin{center}
\includegraphics[scale=0.3]{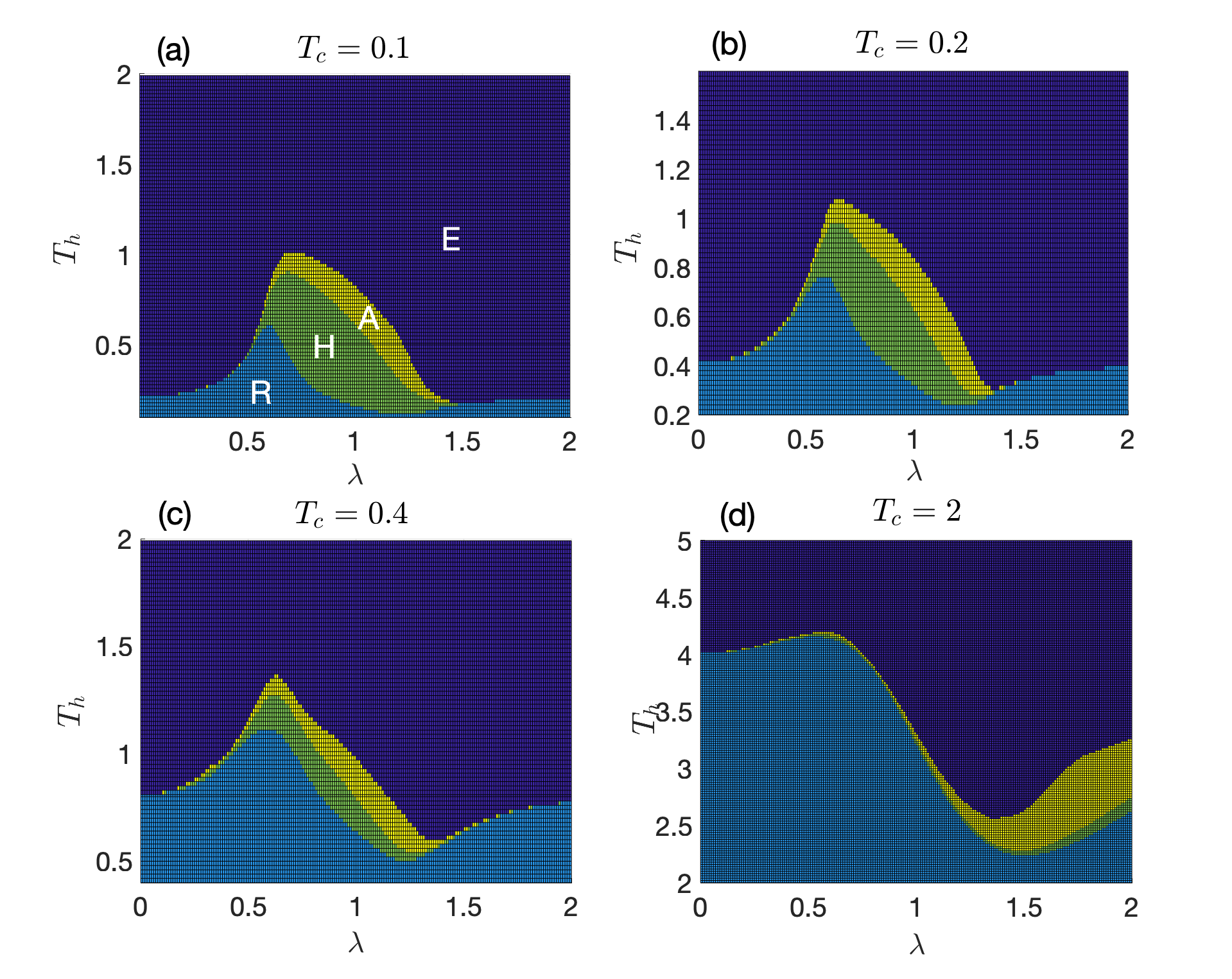}
\end{center}
\caption{The various operating regimes of the quantum Otto machine achieved by varying the temperatures of the hot thermal reservoir $T_{h}$ and the qubit-boson coupling strength $\lambda$, both in units of $\omega$, keeping fixed the temperature of the cold thermal reservoir, also in units of $\omega$ as (a) $T_{c}=0.1$  (b) $T_{c}=0.2$, (c) $T_{c}=0.4$, and (d) $T_{c}=2$. The color code stands for heat engine (magenta), refrigerator (cyan), heater (green), and accelerator (yellow).
The other system parameters are given by $N=8$, $\omega_{h} = 2\omega$, $\omega_{c} = \omega$.}~\label{Fig2}
\end{figure}

\begin{figure}[tbp]
\begin{center}
\includegraphics[scale=0.4]{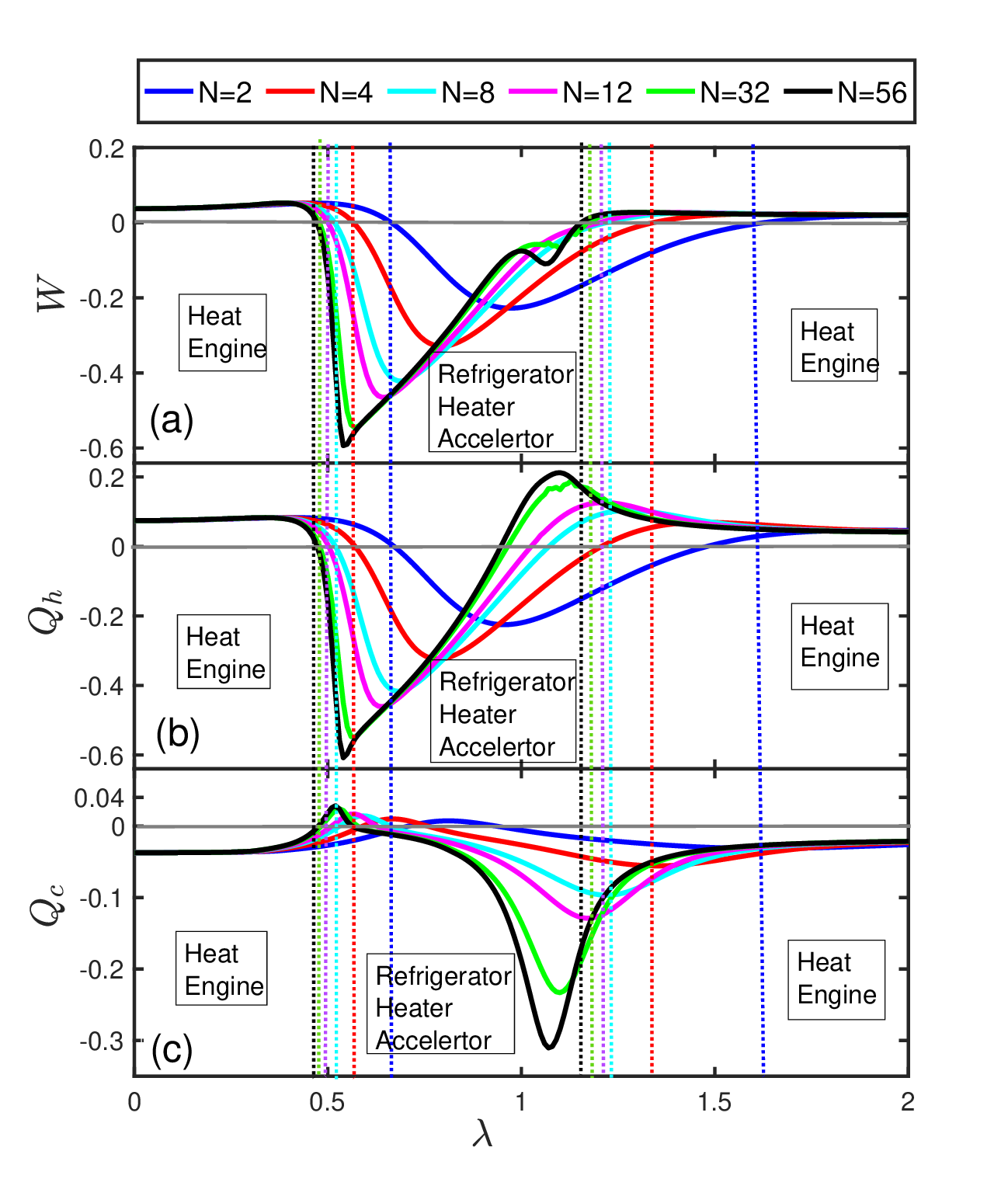}
\end{center}
\caption{(a) Work output $W$, (b) heat $Q_{h}$ and (c) heat $Q_{c}$ as a function of the qubit-boson coupling strength $\lambda$ for different number of qubits $N=2$ (solid blue line), $N=4$ (solid red line), $N=8$ (solid cyan line), $N=12$ (solid magenta line), $N=32$ (solid green  line), $N=56$ (solid black line). Vertical dotted lines divide the different operating regimes of our universal quantum Otto heat machine by the same color used to designate the number N of atoms. The solid horizontal gray line indicates the zero of each quantity.
The other system parameters are given by $T_{h}=0.5$, $T_{c}=0.1$,
$\omega_{h} = 2\omega$, $\omega_{c} = \omega$. All quantities above are in units of $\omega$.
}~\label{fig2-1}
\end{figure}

\begin{figure*}[!htbp]
\begin{center}
\includegraphics[scale=0.4]{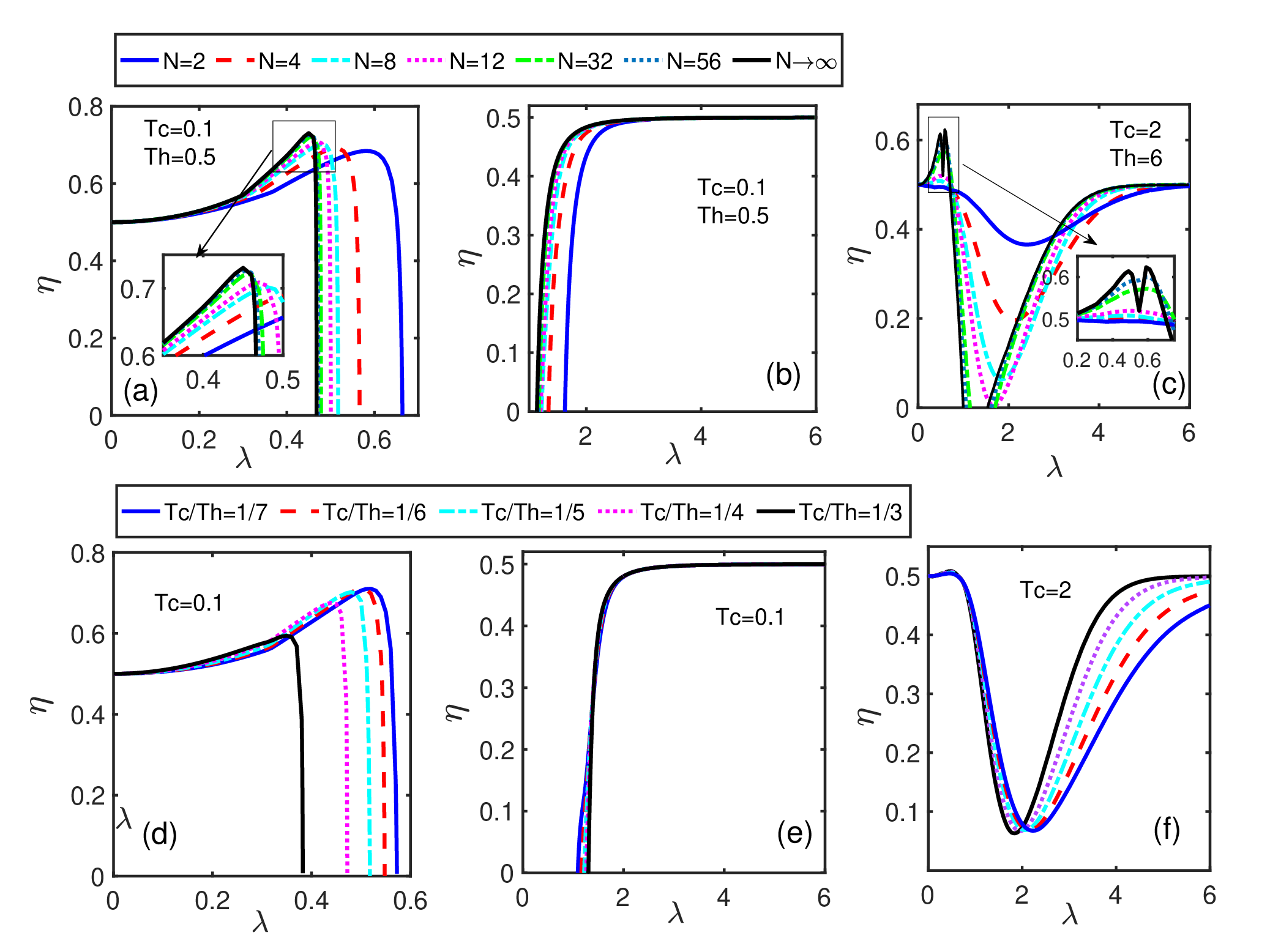}
\end{center}
\caption{ The efficiency $\eta$ of the quantum heat engine as a function of the qubit-boson coupling strength $\lambda$ (in units of $\omega$) for different number of qubits $N=2$ (solid blue line), $N=4$ (dashed red line), $N=8$ (dash dot cyan line), $N=12$ (dotted magenta line), $N=32$ (dash dot  green line), $N=56$ (dotted deep blue line), and infinity N (solid black line), with  fixed $T_{h}=0.5$, $T_{c}=0.1$ for (a) and (b),  $T_{h}=6$, $T_{c}=2$  for (c). Likewise, (d) and (e) and (f) are for the  efficiency $\eta$ of the quantum heat engine as a function of the qubit-boson coupling strength $\lambda$ under different temperature ratios $T_{c}/T_{h}=1/7$ (solid blue line), $T_{c}/T_{h}=1/6$ (dashed red line), $T_{c}/T_{h}=1/5$ (dash dot cyan line), $T_{c}/T_{h}=1/4$ (dotted magenta line), $T_{c}/T_{h}=1/3$ (solid black line), and fixed $N=8$, with cold reservoir temperatures $T_{c}=0.1$ for (d) and (e), and $T_{c}=2$ for (f). Here the temperatures are in units $\omega$. The other system parameters are given by $\omega_{h} = 2\omega$, $\omega_{c} = \omega$.
}~\label{fig23}
\end{figure*}

\begin{figure*}[!htbp]
	\begin{center}
		\includegraphics[scale=0.4]{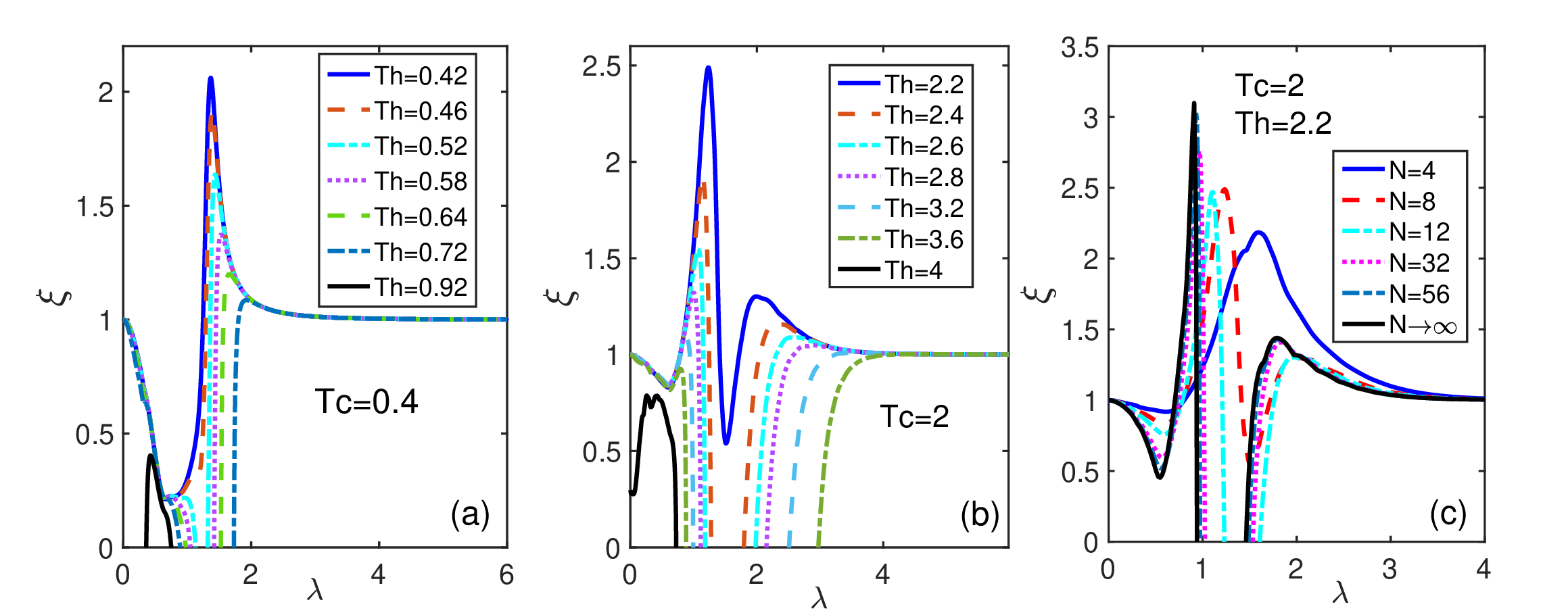}
	\end{center}
	\caption{(a) and (b) show the COP $\xi$ as a function of the qubit-boson coupling strength $\lambda$ (in units of $\omega$)  for different $T_h$ with fixed (a) $T_{c}=0.4$, (b) $T_{c}=2$, and $N=8$ qubits.  (c) COP $\xi$  as a function of the qubit-boson coupling strength $\lambda$ (in units of $\omega$)  under different qubit numbers with fixed $T_{c}=2$, $T_{h}=2.2$. The temperatures are in units of $\omega$.The other parameters of the system are $\omega_{h} = 2\omega$, $\omega_{c}= \omega$.
	}~\label{refri}
\end{figure*}

\begin{figure*}[!htbp]
\begin{center}
\includegraphics[scale=0.4]{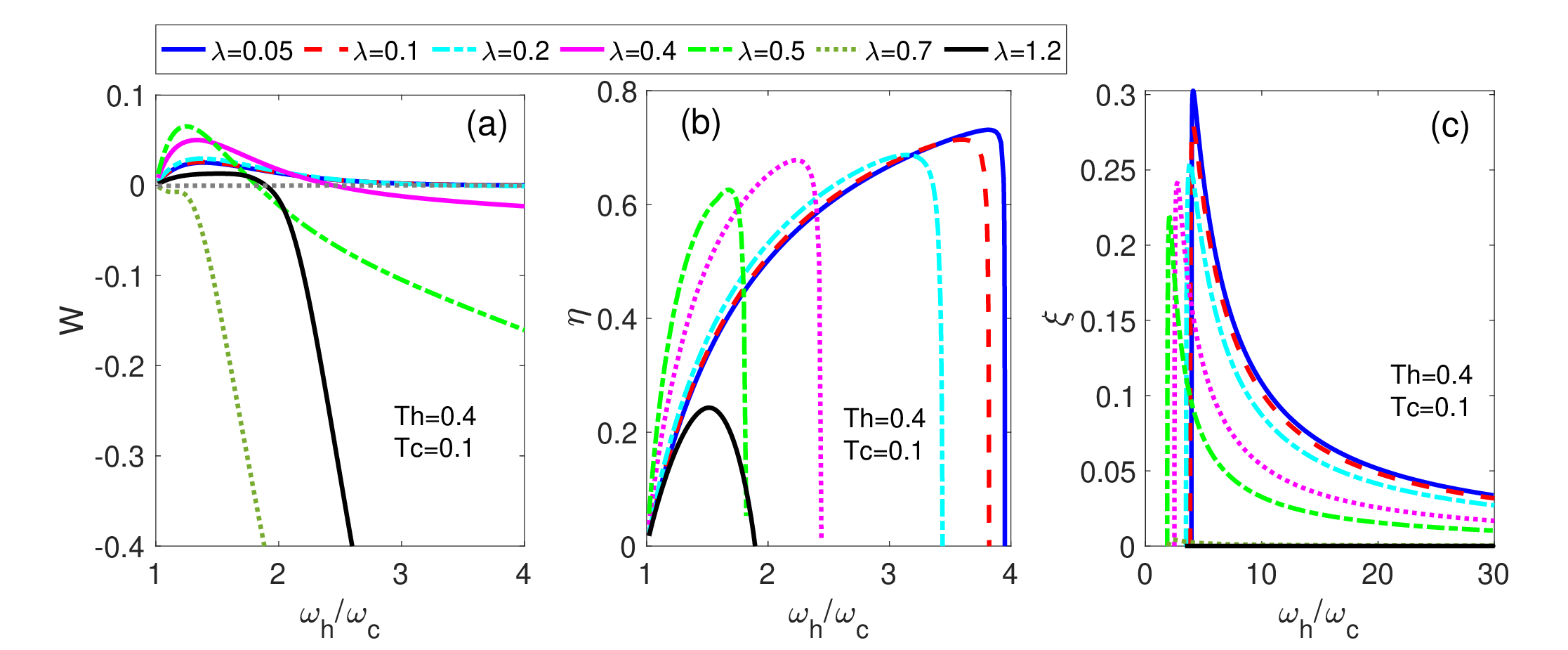}
\end{center}
\caption{(a) Work $W$, (b) efficiency $\eta$, and (c) COP as  a function of the frequency ratio $\omega_{h}/\omega_{c}$ for coupling strength $\lambda=0.05$ (solid blue line), $\lambda=0.1$ (dashed red line), $\lambda=0.2$ (dash dot cyan line), $\lambda=0.4$ (dotted magenta line), $\lambda=0.5$ (dash dot  green line), $\lambda=0.7$ (dash dot  deep green line) and $\lambda=1.2$ (solid black line). The temperatures of the cold and hot reservoirs, given in units of $\omega$, were fixed at $T_{c}=0.1$ and $T_{h}=0.4$, respectively. The number of qubits is fixed to $N=8$.
}
~\label{frequencyratio}
\end{figure*}

\subsection{Working regimes for the universal quantum Otto machine based on the ODM}
We can gain some insight of the Otto cycle by making a qualitative description of the different working regimes for the universal QOHM, as shown in Fig.~\ref{Fig2} for $N=8$ two-level atoms and $\omega_{h}/\omega_{c} = 2$. Note that given two operating temperatures of the Otto cycle, by controlling the parameter $\lambda$ we obtain the four different types of machine. Also, note that the four regions are all present at low temperatures, Fig.2(a)-(c), and mainly occupied by refrigerator (cyan area) and heat engine (magenta area). As the temperature rises, Figs.2(d), the engine and refrigerator operating regions stand out even more. Fig.~\ref{Fig2}(a)-(c) show that heat engine and refrigerator regions are mainly distributed in weak coupling and ultrastrong regime. Strikingly, up to $\lambda  \approx 0.3$ and for $\lambda  \gg 2$, due to the relative harmonicity of the spectrum, the positive-work condition (PWC) follows the one for the quantum harmonic oscillator or qubit, i.e., $T_{h}> \frac{\omega_{h}}{\omega_{c}}T_{c}$.

Next, we analyze the work regimes for the universal QOHM for different numbers N of qubits when the temperatures of the hot and cold thermal reservoirs are fixed. Fig.~\ref{fig2-1} shows the (a) work $W$, (b) heat $Q_{h}$, and (c) heat $Q_{c}$ as a function of the qubit-boson coupling strength $\lambda$ for different values of the qubits number N, thus evidencing the four working regimes for different qubits numbers. As mentioned above, heat engine is distributed in the weak and strong coupling regimes and the deep strong coupling regime, whereas the PWC $T_{h}> \frac{\omega_{h}}{\omega_{c}}T_{c}$ is satisfied, while the refrigerator, heater and accelerator are located around the critical coupling $\lambda_c=\frac{1}{2}\sqrt{\omega_{0}\Delta\coth(\beta\omega_{0}/2)}$, where the spectrum is highly anharmonic. Note that increasing the number of N qubits shifts the engine region to the left allowing engines to be built for smaller values of $\lambda$ as N grows. On the right side, in the deep strong regime, this behavior is maintained: as N grows, the region corresponding to the engine also shifts to the left, increasing the area corresponding to the engine. It is to be noted that for $\lambda \lesssim 0.4$ and for $ \lambda\gtrsim 1.5$ the work extracted by the engine practically remains constant  with the increase of N. This behavior is mimicked by the amounts of heat $Q_{h}$ and $Q_{c}$ exchanged with the reservoirs. Given that increasing the number of atoms means that more work and heat can be exchanged, it is somewhat surprising that for certain values of $\lambda$ increasing N these quantities remain unchanged. On the other hand, for the other types of machines, whose regions correspond to the middle region of Fig.~\ref{fig2-1}, the exchanged work and heat increases with N. 


\subsection{Efficiency and coefficient of performance}

Next, we study the efficiency and COP of the Otto quantum engine and the refrigerator, which are the heat machines of greatest practical interest. To help our analysis, it is useful to recall the analytical result of efficiency $\eta_{\lambda = 0} = 1 - \omega_{c} /\omega_{h}$, and COP  $\xi_{\lambda = 0} =  \omega_{c} /(\omega_{h}-\omega_{c})$, of an Otto cycle when $\lambda = 0$, which corresponds to $N$ qubits and a non-interacting bosonic mode. First, we focus on the engine. In Fig.~\ref{fig23}(a), (b) and (c) we plot efficiency as a function of the qubit-boson coupling strength $\lambda$ for different number of qubits $N=2$ (solid blue), $N=4$ (dashed red), $N=8$ (dash dot cyan), $N=12$ (dotted magenta), $N=32$ (dash dot  green), $N=56$ (dashed dark blue) and the thermodynamic limit $N\rightarrow \infty $ (solid black) with fixed $T_{h}=0.5$, $T_{c}=0.1$ for Figs.~\ref{fig23}(a)-(b), and $T_{h}=6$, $T_{c}=2$ for Fig.~\ref{fig23}(c). Fig.~\ref{fig23}(d) and (e) with $T_c = 0.1$, and Fig.~\ref{fig23}(f) with $T_c = 2$, show the efficiency for various temperature ratios when varying the qubit-boson coupling strength $\lambda$ with fixed $N=8$.
The drop to zero in efficiency, Fig.~\ref{fig23}(a) and (d), and its growth from zero to a maximum, Fig.~\ref{fig23}(b) and (e), occur due to the transition from the engine to the refrigerator regime, corresponding to the regions shown in Fig.~\ref{fig2-1}. But in any case, it is notable that the falls to zero and the rises to the maximum occur suddenly rather than smoothly. Notably, for $\lambda \lesssim 0.3$ the efficiency is independent of the number of atoms used as working substance. Also, note that the efficiency drops to zero for smaller values of $\lambda$ as $N$ grows, as shown in Fig.~\ref{fig23}(a) and (d), because of the shift of the engine region to the left as $N$ grows, as already mentioned when analyzing Fig.~\ref{fig2-1}. 

The main advantage of the ODM over the decoupled system as a working substance happens around the critical region $\lambda_c=\frac{1}{2}\sqrt{\omega_{0}\Delta\coth(\beta\omega_{0}/2)}$, where $\eta> \eta_{\lambda = 0}$ only to $\lambda < \lambda_{c}$, an unexpected result, that means the normal phase of the Dicke model is more suitable to the engine operation than the superradiante phase, as depicted in Figs.~\ref{fig23}(a)-(f). Remarkably, for small temperatures as in Fig.~\ref{fig23}(a) the number of qubits saturates quickly to the thermodynamic limit $N\rightarrow \infty $, being $N \approx 30$ enough to extract the maximum efficiency. Interestingly, for values of temperature ratios for which there is always the engine condition, see Figs.~\ref{fig23}(c) and (f) and also Fig.~\ref{Fig2}, there is still a drop below the $\eta_{\lambda=0}$ around the critical coupling. In the deep-strong coupling regime the efficiency tends to $\eta =0.5$, what is predicted by the effective Hamiltonian Eq.~\eqref{eq11} and the harmonicity of the spectrum for the considered temperatures. We point out that, to small number of atoms $N=8$, there is a decrease in the engine operating region as the temperature gap increases as shows in Fig.~\ref{fig23}(d), the smallest region corresponding to $T_c/T_h = 1/3$ (solid black line). Second, note that the efficiency is smaller the greater the temperature gap is, a somewhat expected behavior when compared with the Carnot efficiency $\eta_{Carnot} = 1 - T_{c} /T_{h}$. As expected, the efficiency of our universal QOHM based on the open Dicke model never exceeds the Carnot efficiency.
From Fig.~\ref{fig23}(a),(b),(d),(e), it would appear that the abrupt drop and sudden resurgence of efficiency values is a characteristic of engine efficiency. However, as can be seen from Fig.~\ref{Fig2}(a)-(d), which shows the region of the various machines, there are temperature ratios for which there will always be a condition for the engine to exist. For these temperature ratios, there will be neither a sudden decrease to zero nor, consequently, an abrupt resurgence in efficiency. In fact, for $T_c>1.0$ in Fig.~\ref{Fig2}(a), $T_c>1.2$ in Fig.~\ref{Fig2}(b), $T_c>1.5$ in Fig.~\ref{Fig2}(c), and $T_c>4.5$ in Fig.~\ref{Fig2}(d) the engine condition will always be fulfilled. This behavior is exemplified in Fig.~\ref{fig23}-(c) and (f), where we explored other temperature ratios for fixed $N =8$ and $T_c=2$. Note from Fig.~\ref{fig23}-(c) that in addition to the abrupt drop in efficiency, which is a signature of the passage from the region that determines engine condition to that of refrigerator, for N=2, 4, and 8, there is also a smooth drop and rise, indicating that despite the increase of $\lambda$ the engine condition continues to be satisfied.

Next, we focus on the refrigerator regime. In Fig.~\ref{refri} the COP $\xi$ as function of the qubit-boson coupling strength is investigated for several ratios of temperatures with fixed $N=8$ and  $T_c=0.4$ - Fig.~\ref{refri}(a), and $T_c=2$ - Fig.~\ref{refri}(b). In Fig.~\ref{refri}(c) we see the effect of the number of qubits on the COP $\xi$. Noteworthy, for the normal phase, for all temperatures and number of qubits, we found $\xi < \xi_{\lambda = 0}$. Besides, similar to the heat engine in the deep-strong coupling regime, the effective Hamiltonian Eq.~\eqref{eq11} leads to an accurate result of $\xi = \xi_{\lambda = 0}$. As evidenced from Fig.~\ref{refri}(a)-(c), the region of coupling $\lambda_c < \lambda \leq 3 $ is where the COP for universal QOHM  having the Dicke model as working substance surpass that of the decoupled system used to fuel the quantum refrigerator. In addition, as observed in Fig.~\ref{refri}(a) and (b), the COP strongly depends on the temperature ratio, thus differing from $\xi_{\lambda = 0} =  \omega_{c} /(\omega_{h}-\omega_{c})$, being higher to small temperature ratios, as the Carnot COP, keeping the limit $\xi \ll \xi_{Carnot}= T_{c} /T_{h}-T_{c}$. We note from Fig.~\ref{refri} that for ratios of temperature where the universal QOHM would work as heat engine with $\lambda=0$ (uncoupled case), corresponding to $\frac{T_{h}}{T_{c}}< \frac{\omega_{h}}{\omega_{c}}$, for some coupling ranges  the universal QOHM  works as a refrigerator with COP lower than that of the uncoupled case.

Lastly, we explore in Fig.~\ref{frequencyratio}(a)-(c) the effect of the frequency ratio $\omega_{h}/\omega_{c}$ on the work exchanged, Fig.~\ref{frequencyratio}(a), as well on the efficiency, Fig.~\ref{frequencyratio}(b), and performance, Fig.~\ref{frequencyratio}(c), for the universal QOHM. The coupling strengths are $\lambda=0.05$ (solid blue line), $\lambda=0.1$ (dashed red line), $\lambda=0.2$ (dash dot cyan line), $\lambda=0.4$ (dotted magenta line), $\lambda=0.5$ (dash dot  green line), $\lambda=0.7$ (dash dot  deep green line) and $\lambda=1.2$ (solid black line). The temperatures were fixed as  $T_{c}=0.1$ to the cold thermal reservoir and $T_{h}=0.4$ to the hot  thermal reservoir. 
The frequency ratios, in addition to indicating more clearly the PWC condition, also allow extracting the point that maximizes both the efficiency and the COP for the universal QOHM. Note the same behavior already observed in other figures, both for efficiency and for COP. In Fig.~\ref{frequencyratio}(b) we see that after a growth until reaching a maximum, there is an abrupt drop, precisely at the point where the engine operating condition changes to the refrigerator condition. In Fig. Fig.~\ref{frequencyratio}(c), where there is a sudden appearance of COP in the refrigerator region, the COP also reaches a maximum and then decreases smoothly. The value of the maximum frequency ratio can thus be used to maximize both efficiency and COP, and for the engine this condition is the so-called  efficiency at maximum power \cite{abah2012single}.

To finish this section, we point out that we also studied two other protocols to carry out the adiabatic processes, namely (i) keeping the frequencies constant and changing the coupling strength and (ii) changing the number of qubits that interact with the quantum mode and fixing both the frequencies and the coupling strength. As verified by our numerical calculations (not shown here), in both protocols the efficiency and the coefficient of performance do not overcome the case in which the working substance is composed of a field mode decoupled from the qubits.


\subsection{Quantum correlations at thermal equilibrium}
In this section, we investigate whether quantum correlations are present at thermal equilibrium and if so, whether they affect the efficiency or COP of the universal quantum heat machine (UQHM) based on the open Dicke model (ODM). In accordance with our analytical and numerical studies, shown in Figs.~\ref{GNl1}(a)-(f), we claim that quantum properties surviving thermalization are not the reason for the superior performance of efficiency, extractable work and COP for the UQHM based on the ODM. The improvements we observed both in efficiency (Fig.5) and COP (Fig.6) are due to the structure of the energy levels, as evidenced by the validity condition $N\gg \langle \hat{b}^{\dagger }\hat{b}\rangle $ to derive the effective Hamiltonian Eq.~\eqref{eq10}, that is, small temperatures require a smaller number of qubits to lead to the anharmonicity around the critical point that is present at all temperatures in the thermodynamic limit.

To calculate quantum correlations, we resort to the following quantities: (i) the second-order correlation function, which captures the occurrence of sub-Poissonian statistics, and (ii) the negativity, which quantifies entanglement. We emphasize that other quantum measures, such as mutual information, squeezing and quantum discord, were investigated and omitted because they showed the same general behavior of the considered quantities.

First, note that the conventional definition of the normalized zero-delay second-order correlation function is~\cite{glauber1963quantum}
\begin{equation}
	g^{(2)}(0) = \frac{\langle (\hat{a}^\dagger)^2(\hat{a})^2\rangle}{\langle \hat{a}^\dagger\hat{a} \rangle^2}.\label{eq:g2-aa}
\end{equation}
This quantity describes the probability of detecting two photons simultaneously. This definition holds for weak light-matter couplings, where the intracavity photons, whose annihilation operator is described by $\hat{a}$, suffice to explain the observed photon correlations. On the other hand, in the USC regime, where the qubit system strongly dresses the bosonic field, the second-order correlation function is derived from the input-output formalism as~\cite{rabl2011photon,ridolfo2012photon}
\begin{equation}
	G^{(2)}(0)=\frac{{\langle}(\hat{X}^-)^2(\hat{X}^+)^2{\rangle}}{{\langle}\hat{X}^-\hat{X}^+{\rangle}^2},~\label{eq:g2-x}
\end{equation}
where
\begin{eqnarray}~\label{xp}
	\hat{X}^+=-i\sum_{k>j}\Delta_{kj}X_{jk}|\phi_j{\rangle}{\langle}\phi_k|,
\end{eqnarray}
with $\hat{X}^-=(\hat{X}^+)^{\dag}$, $\Delta_{kj}=E_k-E_j$ is the energy gap, and $X_{jk}={\langle}\phi_j|(\hat{a}^\dag+\hat{a})|\phi_k{\rangle}$. Here, ${X}_{jk}$ describes the transition from the higher eigenstate $|\phi_k{\rangle}$ to the lower one $|\phi_j{\rangle}$. Notice that, in the weak qubit-photon interaction limit (i.e. $\lambda_i\ll 1$), the operator $\hat{X}^+$ is reduced to $\hat{X}^+=-i\omega\hat{a}$. Thus, the correlation function in Eq.~\eqref{eq:g2-x} simplifies to the conventional case. When $G^{(2)}(0)<1$ the light presents the non-classical effect of anti-bunching, and can be taken as an unequivocal indication of quantumness. 

As for the negativity  $\mathcal{N}(\rho)$ of a subsystem A, it can be defined in terms of a density matrix $\rho$ as ~\cite{eisert1999comparison,zyczkowski1998volume}:
\begin{equation}
	\mathcal{N}(\rho)=\frac{\Vert\rho^{T_{A}}\Vert_{1}-1}{2}
\end{equation}
where $\rho^{T_{A}}$ is the partial transpose of $\rho$ with respect to subsystem A, and $\Vert X\Vert_{1}=Tr|X|=Tr\sqrt{X^{\dagger}X}$ is the trace norm or the sum of the singular value for the operator $X$. Non-zero negativity values indicate the presence of quantum correlations in the form of entanglement, being greater the greater the amount of entanglement present.

\begin{figure}[tbp]
\begin{center}
\includegraphics[scale=0.4]{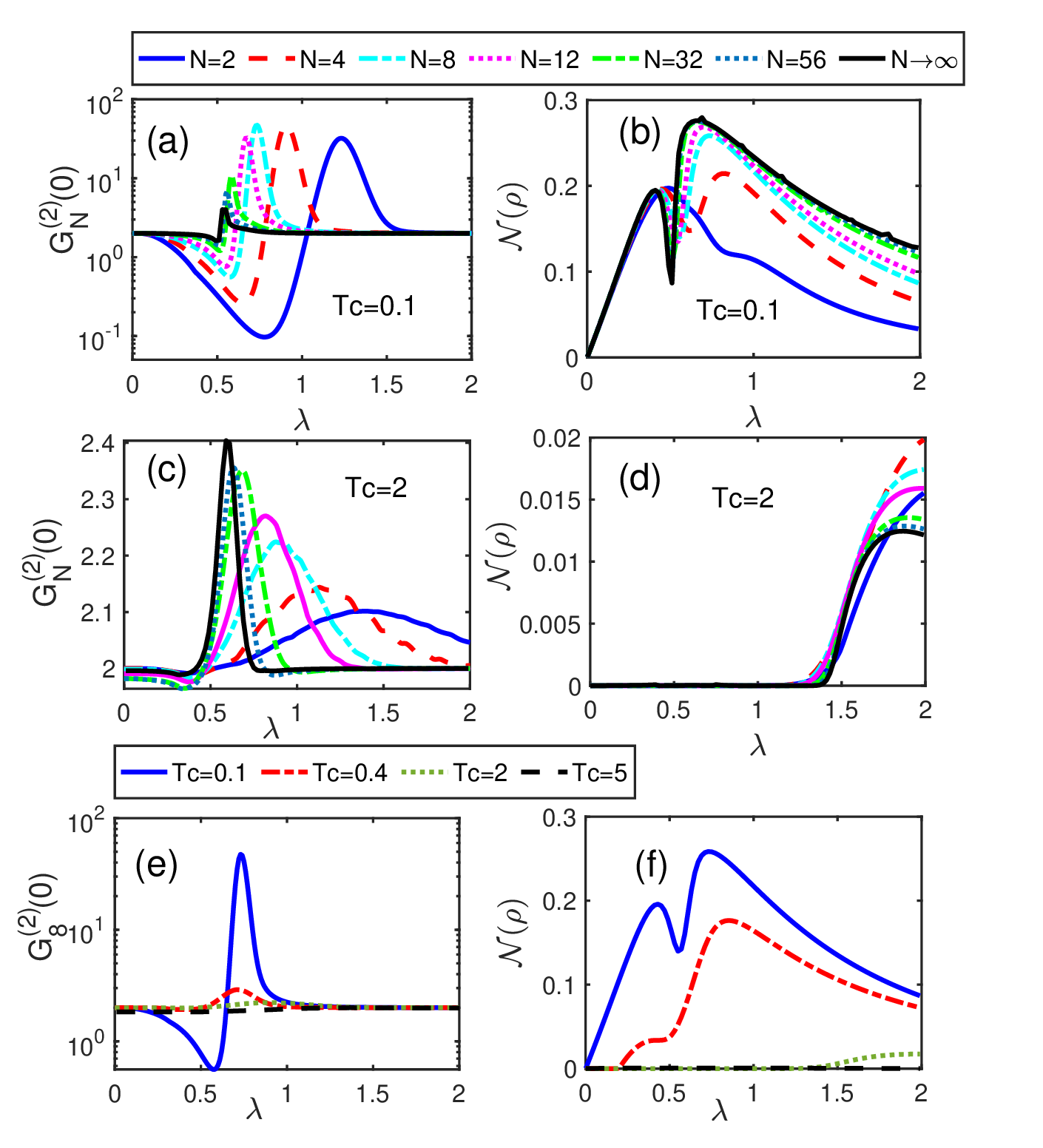}
\end{center}
\caption{Two-photon correlation function $G^{(2)}_N(0)$ (a), (c), (e) and negativity $\mathcal{N}(\rho)$ (b),(d),(f) as a function of the coupling strength $\lambda$.  In (a)-(d) both  the two-photon correlation function and negativity are shown for several qubit number $N$, including the thermodynamic limit, with the cold thermal reservoir temperature fixed at $T_{c}=0.1$  (a)-(b) and  $T_{c}=2$ (c)-(d). In (e)-(f), the qubit number was fixed at $N=8$ and the cold thermal reservoir temperature was chosen as $T_{c}=0.1$ (blue solid line),  $T_{c}=0.4$ (red dash line),  $T_{c}=0.2$ (green dot line), and  $T_{c}=5$ (black dash line). The other system parameters are $\omega_{c} = \omega$. The coupling strength $\lambda$ and temperatures are in units of $\omega$.
}~\label{GNl1}
\end{figure}

We computed $G^{(2)}(0)$ in Figs.~\ref{GNl1}(a),(c),(e) and $\mathcal{N}(\rho)$ in Figs.~\ref{GNl1}(b),(d), (f) as a function of coupling strength for different number of qubits. In Figs.~\ref{GNl1}(e),(f) we set $N =8$ and investigated the effect of different temperatures on the quantumness of the work substance.  
Note that the quantum correlations are degraded by increasing the number $N$ of qubits and increasing the temperature $T_{h}$, whereas $\eta$ and $\xi$ increase with the number of qubits for all temperatures. If we compare the  Figs.~\ref{GNl1}(a)-(f) showing maximum antibunching and maximum entanglement with Figs 4(a)-(f) and 5(a)-(b), which respectively show efficiency and COP, we will see that there is no correspondence between the maximum of quantum correlations and the maximum of efficiency and COP. As far as the second-order correlation is concerned, the efficiency is higher in the deep strong coupling regime ($\lambda >2$), Figs. 4(b),(e), and therefore, far from the region where the second-order correlation shows the sub-Poissonian effect. The same conclusion can be drawn from Fig.5(a)-(c) for the COP, whose maxima lie in regions far from the value of the critical parameter $\lambda$. With regard to negativity, Figs.7(b), (d), and (f) show that the maximum of negativity, and therefore of entanglement, does not coincide with the maximum of efficiency and COP. For example, in Fig. 4(b) for $T_c = 0.1$ and various values of $N$, the efficiency is practically constant with the coupling parameter, and therefore independent of the amount of entanglement, whereas in Fig.7(b) the negativity, also at $T_c = 0.1$ and the various values of $N$ presents maximums and minimums when varying the coupling parameter. The same can be said about the COP: there is nothing in the analysis of the maximums that indicates the relevance of negativity for its improvement. Take for example Fig.7(d) for $T_c =2$, where negativity remains zero for a large range of values of $\lambda$ and then increases monotonically until to $\lambda =2$, with similar behavior even for different values of $N$. Compare with Fig.5(c), where the COP has a very different behavior depending on $N$, with no correspondence with negativity. These same conclusions are supported by additional numerical calculations that we performed (not shown here). To summarize this Section, and as previously mentioned, we point out that the improvement in the efficiency and performance of the UQOHM using the Dicke model as the working substance cannot be attributed to quantum resources ~\cite{de2021efficiency,el2023enhancing}, but it is due to the high anharmonicity of the spectrum around the critical point of the Dicke model.


\section{Conclusion}

In summary, in this work we propose a universal quantum Otto heat machine (UQOHM) based on the open Dicke model (ODM)}. The ODM is composed of N atoms of two levels (qubits) that interact with a mode of the electromagnetic field and both the mode and the N qubits, which constitute the working substance of the universal machine, interact with thermal reservoirs. This model presents a critical point and can be solved analytically in the thermodynamic limit $N\rightarrow \infty $. By universal thermal machine we mean that it is possible, by adjusting the atom-field coupling parameter $\lambda$ of the ODM, to build all types of thermal machines, namely engines, refrigerators, heaters, and accelerators. Focusing on engines and refrigerators, which are the machines with the greatest applicability, we show, for a wide temperature range and a large number of qubits, including in the thermodynamic limit, how the engine efficiency and the refrigerator coefficient of performance change with the parameter $\lambda$ of the ODM. 

We also conducted a study of the quantum correlations present in the ODM using the second order correlation function and negativity showing that, for certain values of the coupling parameter $\lambda$ of the Dicke model, both the antibunching effect and the entanglement survive thermalization. Next, we show that it is possible, close to the critical point, to obtain both an efficiency and a performance for the UQOHM that is greater than the case in which the system is uncoupled, thus showing the advantage of using the Dicke model as the working substance. Furthermore, the detailed study of the second-order correlation function and negativity indicates no correspondence between the improvement in the efficiency and the coefficient of performance of the UQOHM and the quantum resources arising from anti-bunching and entanglement.

\section{Acknowledgement}

We acknowledge financial support from the Brazilian agencies CNPq and FAPEG. This work was performed as part of the Brazilian National Institute of Science and Technology (INCT) for Quantum Information Grant No. 465469/2014-0.  H.-G. X. and J. J. are supported by National Natural Science Foundation of China under Grant No. 11975064.

\newpage
\bibliographystyle{unsrt}
\bibliography{bibliography.bib}

\end{document}